\documentclass[english,aps,prx,twocolumn,superscriptaddress,longbibliography]{revtex4-2}
\usepackage{amssymb}
\usepackage{amsmath}
\usepackage{graphicx}
\usepackage{natbib}
\usepackage{epstopdf}
\usepackage{color}
\usepackage{braket}
\usepackage{physics}
\usepackage{mathrsfs}
\usepackage{upgreek}
\usepackage{todonotes}
\usepackage[colorlinks=true, allcolors=blue]{hyperref}
\begin{document}
%\title{AFM-based Charge-Locking in Silicon Quantum Devices} %not sure the average reader will understand the significance of this title
%Newer options
%\title{Second Quantization: Gating a Few Electron Quantum Dot Through the Discrete Charging of a Nanoscale Floating Gate}
\title{Second Quantization: Gating a Quantum Dot Through the Sequential Removal of Single Electrons from a Nanoscale Floating Gate}
%\title{Second Quantization: Gating a Few Electron Quantum Dot Through the Sequential Removal of Single Electrons from a Nanoscale Floating Gate}
%\title{Second Quantization: Gating a few electron quantum dot through the discrete single electron charging of a nanoscale floating gate}

\author{Artem O. Denisov}
\email{adenisov@princeton.edu}
\affiliation{Department of Physics, Princeton University, Princeton, New Jersey 08544, USA}
\author{Gordian Fuchs}
\affiliation{Department of Physics, Princeton University, Princeton, New Jersey 08544, USA}
\author{Seong W. Oh}
\altaffiliation{Present Address: Department of Electrical and Systems Engineering, University of Pennsylvania, Philadelphia, Pennsylvania~19104, USA}
\affiliation{Department of Physics, Princeton University, Princeton, New Jersey 08544, USA}
\author{Jason R. Petta}
\email{petta@physics.ucla.edu}
\affiliation{Department of Physics, Princeton University, Princeton, New Jersey 08544, USA}
\affiliation{Department of Physics and Astronomy, University of California, Los Angeles, California 90095, USA}
\affiliation{Center for Quantum Science and Engineering, University of California, Los Angeles, California 90095, USA}

\begin{abstract}
We use the tip of an atomic force microscope (AFM) to charge floating metallic gates defined on the surface of a Si/SiGe heterostructure. The AFM tip serves as an ideal and movable cryogenic switch, allowing us to bias a floating gate to a specific voltage and then lock the charge on the gate by withdrawing the tip. Biasing with an AFM tip allows us to reduce the size of a quantum dot floating gate electrode down to $\sim100~\mathrm{nm}$. Measurements of the conductance through a quantum dot formed beneath the floating gate indicate that its charge changes in discrete steps. From the statistics of the single-electron leakage events, we determine the floating gate leakage resistance $R \sim 10^{19}~ \mathrm{Ohm}$ - a value immeasurable by conventional means. 
\end{abstract}

\maketitle

\section{Introduction}
Spin qubits in gate-defined quantum dots (QDs) have recently demonstrated two qubit gates with fidelities exceeding fault-tolerant thresholds~\cite{doi:10.1126/sciadv.abn5130, Xue2022, Noiri2022}. Spin qubits have also been integrated with microwave cavities, enabling dispersive spin state readout, coherent spin-photon coupling, and microwave-mediated spin-spin interactions~\cite{Mi2018, Landig2018,doi:10.1126/science.aar4054,Borjans2020,PhysRevX.12.021026}. Long coherence times~\cite{Tyryshkin2012}, compatibility with industrial fabrication techniques~\cite{Maurand2016, Zwerver2022}, and the potential for integration with classical silicon electronics~\cite{Xue2021} make them one of the most promising platforms for quantum computing~\cite{PhysRevA.57.120,2112.08863}. 

As quantum processors increase in complexity, many solid state qubit platforms will soon face the challenge of delivering a growing number of room temperature control signals to the cryogenic environment of the chip. Since most quantum error correction protocols require qubits to be arranged in two-dimensional (2D) arrays~\cite{PhysRevA.86.032324}, qubit interconnect crowding is an outstanding issue~\cite{2209.06609}. For semiconductor spin qubits, where qubits are only separated by $\sim$ 100 nm, the interconnect challenge is further exacerbated~\cite{doi:10.1063/1.4922249, PhysRevApplied.6.054013}. 

\begin{figure*}[tbh!]
	\includegraphics[width=1.9\columnwidth]{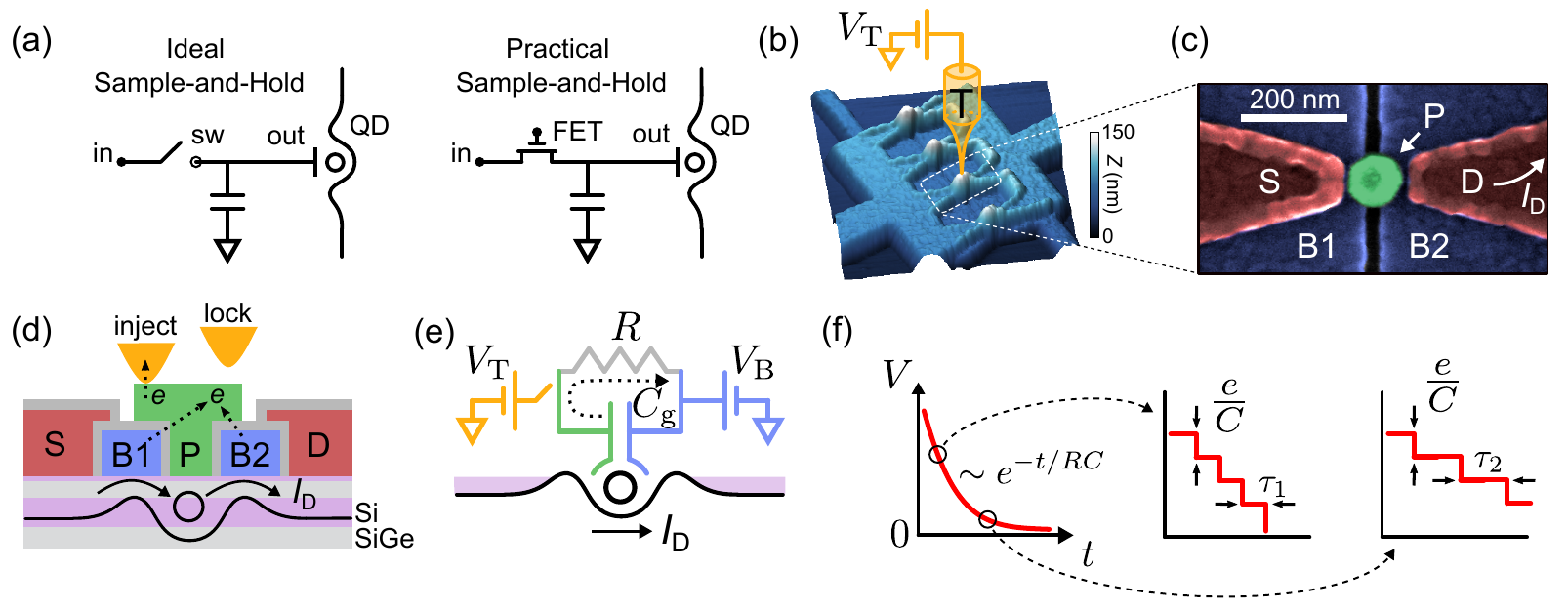}
	\caption{(a)~Ideal and practical sample-and-hold circuits for locking charge on a QD gate electrode. (b)~Low-temperature topographic AFM image of the device, which consists of four identical sections, each containing a floating metallic gate. The dc-biased tip (T) is used to charge a metallic floating gate. (c)~False-color scanning electron microscope image of a section of the device, with Al source-drain (S/D) accumulation gates, Al barrier gates (B1/B2), and a floating Pd gate P. $I_{\rm D}$ is the current through the charge detector QD formed beneath the P-gate. (d)~Cross-section of the device. The tip is used as a switch to inject and lock the charge on the floating gate. (e)~Equivalent circuit. Colors correspond to the sketch in~(d): switch/tip (yellow), resistance/gate-oxide layer (gray), barrier gates (blue), P-floating gate (green), and QD induced in the Si quantum well (black). (f)~$RC$ discharge curve in the quantum limit allows the independent extraction of the capacitance $C$ and resistance $R$ by utilizing the quantization of electric charge. $R$ is related to the average waiting time $\langle \tau \rangle$ between two uncorrelated tunneling events as $I=e/\langle \tau \rangle=V/R$, while $C$ is determined from the size of the discrete voltage steps $e/C$.}
	\label{fig_1}
\end{figure*}

A primary approach that is being pursued to reduce the number of room temperature control lines is charge-locking~\cite{Vandersypen2017,Veldhorst2017,1912.01299}. Charge locking is conceptually related to classical dynamic random access memory~\cite{Keeth2007DRAMCD}. In the case of gate-defined QDs, the concept is to electrically detach the QD gate electrode from the control line. If the electrical isolation of the gate is high, it will retain its charge for a sufficiently long time, allowing other gates on the device to be manipulated with the same room temperature control line. In the perfect case, the so-called ``sample-and-hold" circuit~\cite{Horowitz:1981307}, a zero-resistance switch is desired to dynamically charge and lock the floating node as shown in Fig.~\ref{fig_1}(a). Practically it is quite challenging to realize a mechanical switch for mesoscopic devices. To date only field-effect transistors (FETs), integrated both on-chip and off-chip, have been used to isolate gates from room temperature signals in GaAs~\cite{1912.01299, doi:10.1063/1.4932012}, Si/SiGe~\cite{doi:10.1063/1.4807768, doi:10.1063/5.0012883}, and CMOS~\cite{PhysRevApplied.9.054016, Schaal2019} QDs.

Interconnect crowding in 2D qubit arrays can be significantly eased by vertically bringing electrical connections out of the device plane. Air bridges~\cite{doi:10.1063/1.4863745} and thru-silicon vias with indium bumps have been successfully implemented for superconducting qubits~\cite{Rosenberg2017, 2103.08536}. Flip-chip bonding is being pursued for cavity-coupled Si spin qubits~\cite{Holman2021}. A CMOS-compatible device architecture incorporating nanoscale vias was recently introduced~\cite{doi:10.1021/acs.nanolett.1c03026}, but its application has so far been limited to small linear arrays of Si/SiGe QDs~\cite{HRL_2QEO}.

Inspired by the potential of via-based 3D integration~\cite{doi:10.1021/acs.nanolett.1c03026, HRL_2QEO}, we explore the nanoscale limit of the floating gate approach by fabricating isolated gates with diameter $\sim100~\mathrm{nm}$. We attempt to realize the ideal sample-and-hold circuit by using the tip of an atomic force microscope (AFM) as a movable voltage node~\cite{Seong_AIP,doi:10.1021/acs.nanolett.2c01098}. The floating gate can then be electrically charged by contacting it with a voltage-biased AFM tip. Withdrawing the AFM tip locks the charge on the floating gate. We probe the charge retention of the floating gate by measuring the conductance through a QD defined in a Si quantum well beneath the floating gate. The true floating gate design and its small $\sim$ 100 nm diameter eliminates problems typical for FET-based charge-locking, such as charge leaking through the FET channel due to a finite on-off ratio~\cite{PhysRevApplied.9.054016}.

We first demonstrate that the AFM tip can be used as a stationary vertical gate while in continuous electrical contact with the metallic floating gate. By sweeping the dc tip voltage, $V_{\rm T}$, we can tune the QD charge occupancy as in a conventional QD device. Then, by withdrawing the biased tip, we show that the electric charge is locked on the floating gate over a time-scale set by the leakage resistance $R$ to other gates on the device. We confirm the non-destructive character of the injection/locking process by repeating it multiple times on a single gate and obtaining a reproducible charge-sensor response down to individual electron tunneling events.

The nanoscale dimensions of our floating gate allows us to explore an intriguing ``second quantized'' operating regime since the voltage resolution of the gate is fundamentally limited by the quantization of electric charge~\cite{doi:10.1126/science.275.5300.649}. Due to a sub-fF total capacitance $C_{\mathrm{g}}$ of the gate, only a few hundred electrons have to be removed from the gate to induce a few-electron QD beneath it. In addition, we directly probe the retention time of the charge locked on the floating gate. We utilize the QD induced in the quantum well under the floating gate as a highly sensitive charge sensor~\cite{Lu2003,PhysRevLett.96.076605} to extract values of $C_{\mathrm{g}}$ and $R$. In devices with a single layer of gates, we show that charge can stay locked for several hours.

\begin{figure*}[tbh!]
	\includegraphics[width=1.6\columnwidth]{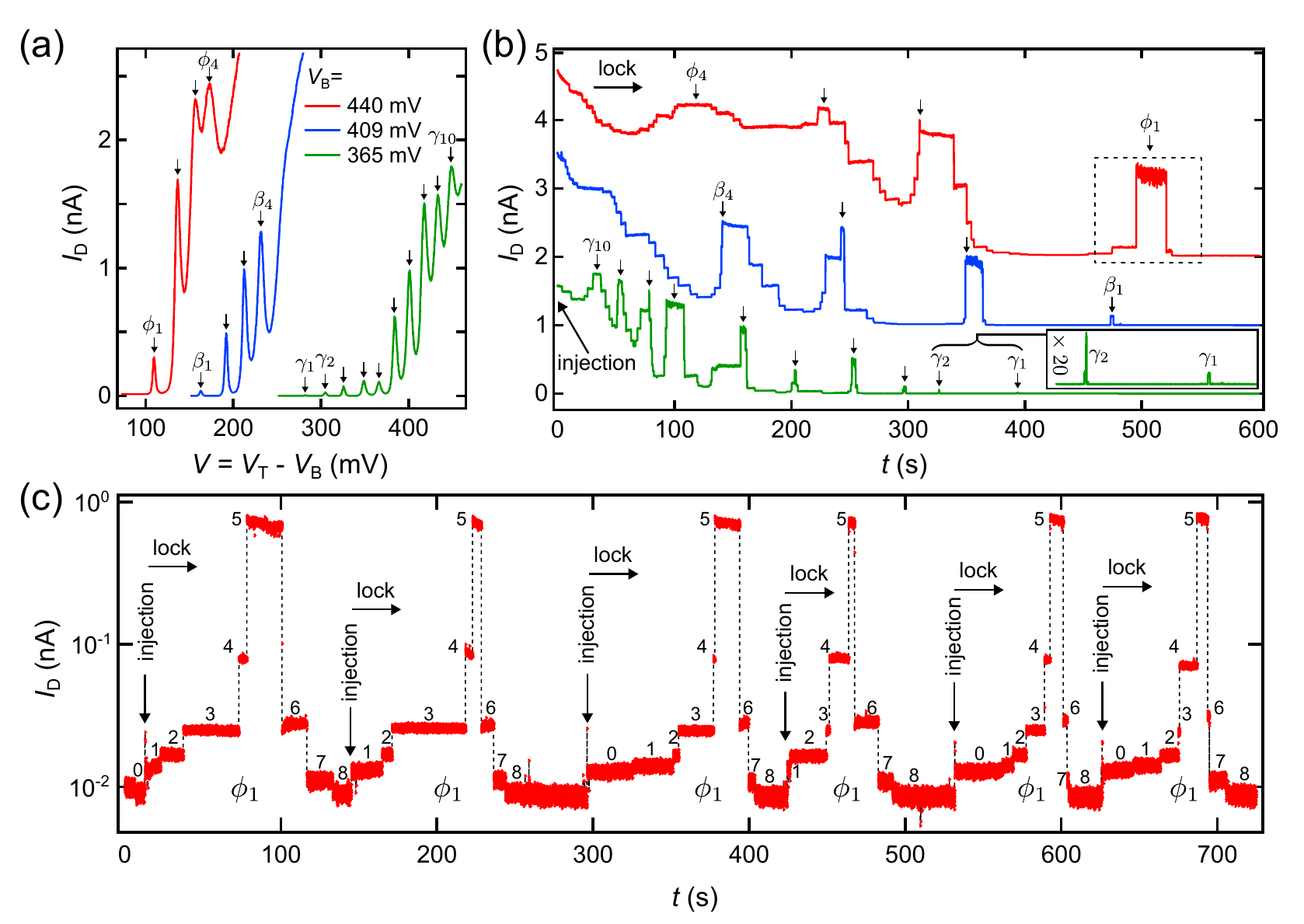}
	\caption{(a)~Continuous injection. Coulomb blockade peaks (CBPs) are measured as a function of the differential voltage $V$ = $V_{\rm T} - V_{\rm B}$ while the AFM tip continuously touches the floating gate. The three traces are taken at different barrier voltages ($V_{\rm B}$ = 365, 409, and 440 mV) to cover a wider range of differential voltage. For each curve, the series of visible CBPs is marked as $\phi_{\mathrm{i}}$, $\beta_{\mathrm{i}}$, and $\gamma_{\mathrm{i}}$. (b)~Charge locking. Current measured as a function of time after the biased tip is moved away from the floating gate. Each step corresponds to a single electron tunneling from the floating gate to a barrier gate. The shadings indicate the averaging interval between two CBPs as in (a). (c)~Reproducibility of the discrete current steps around Coulomb peak $\phi_{\mathrm{1}}$ over six repetitive injection cycles. The number of electrons leaked after injection is indicated for each cycle.}
	\label{fig_2}
\end{figure*}

\section{Experimental setup}
Our accumulation-mode device is fabricated on an undoped Si/SiGe heterostructure consisting of a $5~\mathrm{nm}$ thick Si quantum well (QW) that is buried under a $50~\mathrm{nm}$ thick layer of Si$_{\mathrm{0.7}}$Ge$_{\mathrm{0.3}}$ and a $2~\mathrm{nm}$ thick Si cap. To form a QD in the plane of the QW, we utilize a gate stack consisting of three overlapping layers as shown in Figs.~\ref{fig_1}(c,d). Two Al layers form barrier (B1, B2) and accumulation (S, D) gates, while the Pd disc on top serves as a floating plunger gate. Pd is used for the final layer as it enables good electric contact to the AFM tip~\cite{PhysRevLett.100.197002}. In this work we focus on a small section of the device, indicated by the white dashed line in Fig.~\ref{fig_1}(b), while other sections of the device show similar behavior. 

During charge injection [Fig.~\ref{fig_1}(d)], the tip stays in contact with the floating gate and $V_\mathrm{T}=V_\mathrm{P}$. Charge-locking takes place when the biased tip is lifted away from the device ($\sim200~\mathrm{nm}$ above the gate). After the AFM tip is extracted, the charge trapped on the floating gate can leak to the Si/SiGe substrate or the Al barrier gates (B1, B2). We measure the current $I_{\mathrm{D}}$ through the QD induced beneath the P-gate to sense the real-time dynamics of the locked charge. The equivalent circuit is shown in Fig.~\ref{fig_1}(e) where the switch denotes the tip, while the tunnel junction between the floating gate and barriers is depicted as an $RC$ circuit capacitively coupled to the Si/SiGe QD.

In a classical discharging circuit, the $RC$ time constant can be extracted from the exponential decay of voltage $V$ across the capacitor as shown in the left panel of Fig.~\ref{fig_1}(f). However, the individual values of $R$ and $C$ remain unknown. Nevertheless, $R$ and $C$ can be extracted independently if the electrometer is sensitive and fast enough to resolve single electron tunneling events, as sketched in Fig.~\ref{fig_1}(f)~\cite{GUSTAVSSON2009191}. The capacitance is directly related to the size of the discrete voltage steps $e/C$. At the same time the resistance can be extracted from the electric current $I=e/\langle \tau \rangle=V/R$, where $\langle \tau \rangle$ is the average waiting time before the next tunneling event. Note that in this model $\langle \tau \rangle$ increases linearly as $V$ decreases. %In our system, the QD charge detector is sensitive enough to fulfill the condition that the width of the detector Coulomb blockade peaks is comparable to the size of the voltage step: $\mathrm{max}\{kT,eV_{\mathrm{bias}}\}/\alpha\sim e/C_{\mathrm{g}}$, where $T$ is the electron temperature, $\alpha$ is the lever arm of the floating gate, $C_{\mathrm{g}}$ is the total capacitance of the floating gate, and $V_{\mathrm{bias}}$ is the bias voltage between charge detector source and drain ohmic contacts.

\section{AFM-based charge-locking}
We begin by measuring Coulomb peaks with the tip in contact with the floating gate, such that $V_{\mathrm{T}}$ = $V_{\mathrm{P}}$. Figure~\ref{fig_2}(a) shows $I_{\rm D}$ through the QD induced in the QW as a function of the voltage difference between the floating gate and barrier gates $V~=~V_{\mathrm{T}}-V_{\mathrm{B}}$. For these experiments the barrier gates are kept at the same potential $V_{\mathrm{B}}$ = $V_{\mathrm{B1}}$ = $V_{\mathrm{B2}}$. Three traces are acquired with different $V_{\mathrm{B}}$ to cover a broader range of $V$ and $\langle\tau\rangle$. We mark each visible Coulomb peak: $\phi_{\mathrm{i}}$, $\beta_{\mathrm{i}}$ and $\gamma_{\mathrm{i}}$ for the red, blue, and green curves respectively.

By suddenly lifting the tip from the floating gate and monitoring $I_{\rm D}$, we study the retention of the locked charge for three values of $V_{\mathrm{B}}$ as shown in Fig.~\ref{fig_2}(b). These data sets are color-matched to a corresponding curve in Fig.~\ref{fig_2}(a). Just before the tip was lifted around $t=0~\mathrm{s}$ (injection), $V$ for each curve was slightly exceeding the last prominent CBP in Fig.~\ref{fig_2}(a) ($\phi_{\mathrm{4}}$, $\beta_{\mathrm{4}}$ and $\gamma_{\mathrm{10}}$). As time passes, the floating gate discharges and $I_{\mathrm{D}}$ retraces the series of Coulomb peaks shown in Fig.~\ref{fig_2}(a). In contrast to continuous injection, where the AFM tip is in contact with the floating gate, discharge occurs in discrete steps. In Fig.~\ref{fig_2}(c), we show that single-electron tunneling is the dominant process here and the jumps are solely related to the electrostatic environment around the QD and are not just random measurements artifacts due to $1/f$ noise. Here we sequentially repeat injection and locking of charge around the CBP $\phi_{1}$, as highlighted by the dashed square in Fig.~\ref{fig_2}(b). After each injection, the current $I_{\mathrm{D}}$ evolves through the same sequence of discrete steps marked from 0 to 8, the number of additional electrons that have tunneled to the floating gate. The reproducibility of the plateaus in $I_{\rm D}$ implies that the rates of all higher-order tunneling processes are much slower. Note that we work at differential voltages that are much larger than the superconducting gap $\Delta\approx200~\mu\mathrm{eV}$ of Al. However, very rarely [e.g. for 2 out of 50 events in Fig.~\ref{fig_2}(c)], the waiting time between two tunneling events appears to be shorter than the integration time ($\sim10~\mathrm{ms}$) making these events indistinguishable from two-electron tunneling. As a result, the measured total capacitance $C_{\mathrm{g}}$ of the floating gate, which can be extracted directly from the time traces~\cite{doi:10.1126/science.285.5434.1706} as described below, acquires imprecision due to these counting errors.

\section{Capacitance measurement by counting electrons}
\begin{figure}[tbh!]
	\includegraphics[width=1\columnwidth]{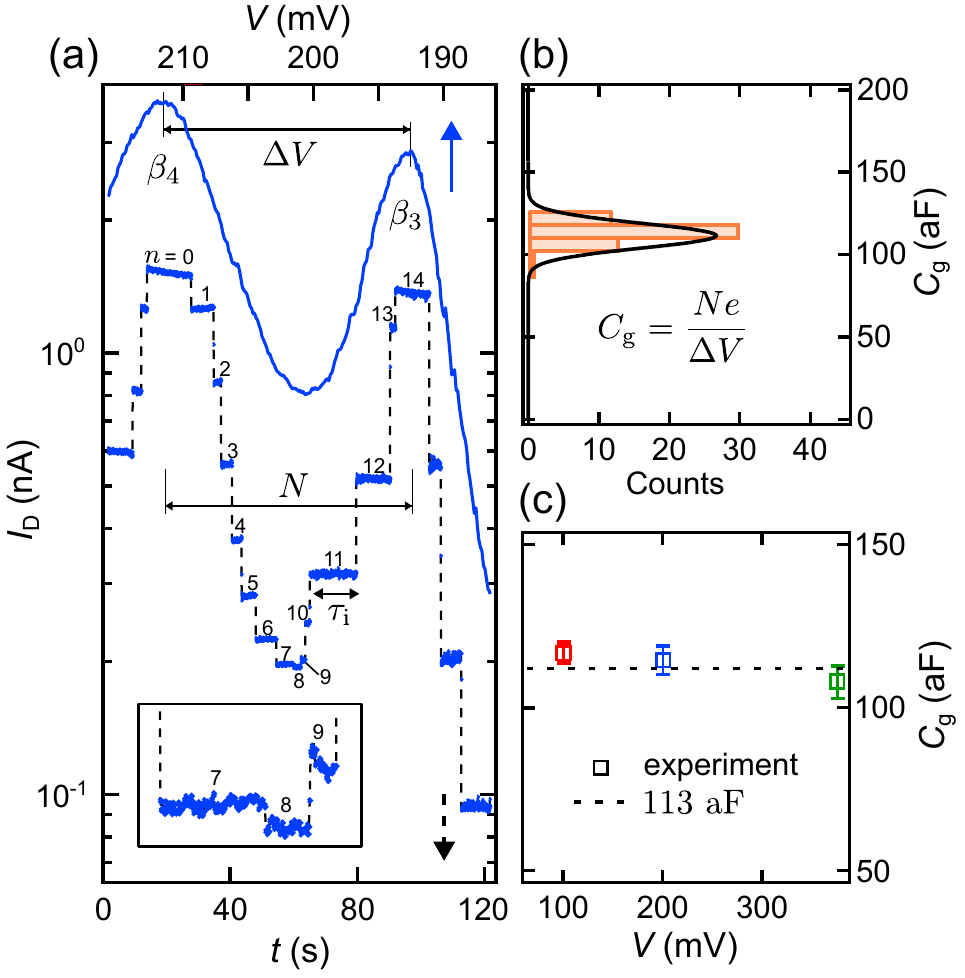}
	\caption{(a)~Current $I_{\rm D}$ between the $\beta_{3}$ and $\beta_{4}$ CBPs plotted as a function of the differential voltage (solid, top axis) and time after injection (dashed, bottom axis). The solid data set is vertically offset by $1~{\mathrm{nA}}$ for clarity. The number of tunneling events starting from the local current maximum is numbered by $n$. The total number of electrons to cover the voltage difference $\Delta V$ between two CBPs is denoted as $N$. The inset shows enlarged data in the dip between Coulomb peaks. (b)~The histogram shows the distribution of the floating gate capacitance $C_{\mathrm{g}}$ due to random errors over all collected data sets. (c)~Averaged $C_{\mathrm{g}}$ as a function of differential voltage.}
	\label{fig_3}
\end{figure}

In Fig.~\ref{fig_3}(a) we plot side-by-side the pair of CBPs $\beta_3$ and $\beta_4$ [marked by blue shading in Figs.~\ref{fig_2}(a, b)] as a function of $V=V_{\mathrm{T}}-V_{\mathrm{B}}$ (top axis) and time (bottom axis). For clarity, the upper data set is offset by 1 nA. The voltage difference between the peaks $\Delta V$ can be alternatively covered by $N$ electrons tunneling to the floating gate such that $N-1<\Delta V C_{\mathrm{g}}/e<N$. By definition the total capacitance can be measured through single electron counting. We repeat the injection-locking cycle multiple times for three different differential voltages between $\phi_1$ and $\phi_2$, $\gamma_6$ and $\gamma_7$, and $\beta_3$ and $\beta_4$ from Figs.~\ref{fig_2}(a, b) and plot the histogram of measured $C_{\mathrm{g}}$'s in Fig.~\ref{fig_3}(b). The narrow distribution implies a single dominant tunneling process, which is slightly widened towards lower capacitance by two-electron processes and towards higher capacitance by large $1/f$ charge noise events that are interpreted as electron tunneling events.

The extracted $C_{\mathrm{g}}=113\pm4~\mathrm{aF}$ is plotted as a function of differential voltage in Fig.~\ref{fig_3}(c). As expected, $C_{\mathrm{g}}$ shows no dependence on $V$ since the capacitance is determined only by the device geometry. The error in $C_{\mathrm{g}}$ originates from both the confidence interval of the mean value in Fig.~\ref{fig_3}(b) and from the systematic one-electron uncertainty between $N-1$ and $N$. It should be noted that the measured value is the total capacitance of the floating gate, which can be further broken into the sum of tunnel junction (TJ) and floating gate-to-QD capacitances: $C_{\mathrm{g}}=C_{\mathrm{TJ}}+C_{\mathrm{QD}}$. The latter can be estimated simply as $C_{\mathrm{QD}}=e/\Delta V\approx 10~\mathrm{aF}$.

%\textcolor{red}{This uncertainty can be potentially eliminated by analyzing the height difference between current plateaus which is beyond the scope of this work.} 

AFM-based charge locking allows us to reduce the footprint and stray capacitance $C_{\mathrm{g}}~=~113~\mathrm{aF}$ of the floating node by a factor of 700 -- 7000 compared to previous FET-based experiments~\cite{PhysRevApplied.9.054016, doi:10.1063/5.0012883}. From a metrological perspective, our approach allows a direct measurement of the sub-fF total capacitance of an isolated object by counting electrons using a QD charge sensor. In previous work~\cite{doi:10.1126/science.285.5434.1706}, the capacitance standard based on counting electrons was 16,000$\times$ larger.

\section{Tunneling statistics and resistance}
\begin{figure*}[tbh!]
	\includegraphics[width=1.9\columnwidth]{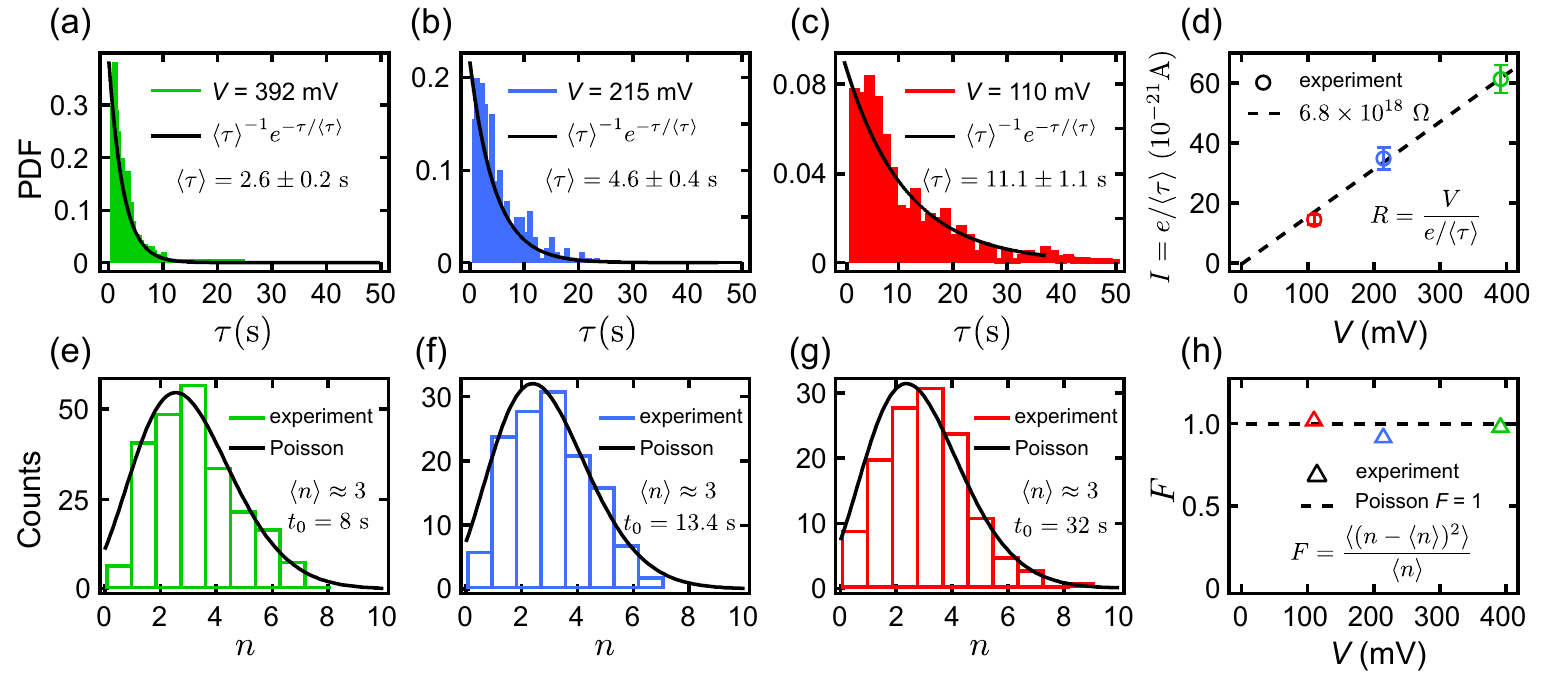}
	\caption{(a-c)~The probability density function of the wait time distribution measured between consecutive tunneling events for various differential voltages. The solid lines correspond to exponential distribution fits with a mean value of $\langle\tau\rangle$. (d) The junction resistance can be extracted from a plot of the current $I=e/\langle\tau\rangle$ as a function of differential voltage. The dashed curve corresponds to $R=6.8\times10^{18}~\Omega$. (e-g)~Statistical distribution of the number $n=3$ of electrons tunneling to the floating gate during a given time $t_{0}$. Three panels and their colors correspond to three differential voltages. The time $t_{0}$ is chosen to have the same mean number of events $\langle n\rangle\approx 3$. The solid lines show Poissonian fits with expected occurrence rate of $\lambda=t_{0}/\langle\tau\rangle$. (h)~Fano factor plotted as a function of the differential voltage. The dashed line corresponds to the expected value $F=1$ for an uncorrelated Poisson process.}
	\label{fig_4}
\end{figure*}
From time traces similar to the one shown in Fig.~\ref{fig_3}(a), we can directly extract the statistical properties of sequential electron transport to the floating gate~\cite{PhysRevLett.96.076605} and confirm it's uncorrelated nature. In Figs.~\ref{fig_4}(a--c), we plot probability density functions (PDF) of the waiting time distribution between adjacent tunneling events at various differential voltages. As expected for the uncorrelated transport of particles through the highly non-transparent TJ \cite{PhysRevLett.75.1610}, the waiting times are distributed exponentially~\cite{BLANTER20001} ${P}(\tau)={\langle\tau\rangle}^{-1} e^{-\tau/\langle\tau\rangle}$, where $\langle\tau\rangle$ is the mean time interval between tunneling events. Additionally, we can extract the time correlation transport properties as shown in Figs.~\ref{fig_4}(e--g). Here we plot the distribution of the number $n$ of events during a given time window $t_{0}$, which is chosen to fit roughly the same average number of events $\langle n \rangle\approx3$. We checked that this choice does not affect the results. The theoretical Poisson distributions (solid lines) $P(n)=\lambda^{n}e^{-\lambda}/n!$, where the occurrence rate $\lambda=t_{0}/\langle \tau\rangle$ is determined experimentally by the mean time, match the experimental data very well, given that no fitting parameters are used. The second central moment (shot-noise) of the distribution $F={\langle(n-\langle n\rangle)^{2}\rangle}/{\langle n\rangle}$, known as the Fano factor~\cite{BLANTER20001} closely fits the tunnel junction limit $F=1$~\cite{PhysRevLett.75.1610, doi:10.1126/science.1084647} as shown in Fig.~\ref{fig_4}(h).

The extracted mean time interval $\langle\tau\rangle$ between tunneling events can be converted~\cite{PhysRevB.72.115331} to the average electrical current $I=e/\langle \tau\rangle$, plotted in Fig.~\ref{fig_4}(d) as a function of differential voltage. The resulting $I-V$ curve originates at the origin, confirming the statement that the charge leaks solely through the barrier gates. From the linear fit, we can extract the resistance of the TJ: $R=6.8\times10^{18}~\Omega$. Such a high value is immeasurable by conventional means and the FET-based charge locking technique. The latter is because even a lower estimate of the typical FET stray capacitance of $C_{\mathrm{stray}}\sim100~\mathrm{fF}$~\cite{doi:10.1063/5.0012883, Schaal2019} i) results in an almost infinite time constant $RC_{\mathrm{stray}}\approx700,000~\mathrm{s}\approx 8~\mathrm{days}$ (taking the data for Fig.~\ref{fig_4}(d) would have taken months) and ii) the QD sensor must be highly sensitive to catch voltage jumps of $e/C_{\mathrm{stray}}\approx1.5~\mu V$. We addressed this limitation by dramatically reducing the stray capacitance of the floating gate (1000 times) to achieve a feasible time constant of $\sim700~\mathrm{s}$.

\begin{figure}[tbh!]
	\includegraphics[width=1\columnwidth]{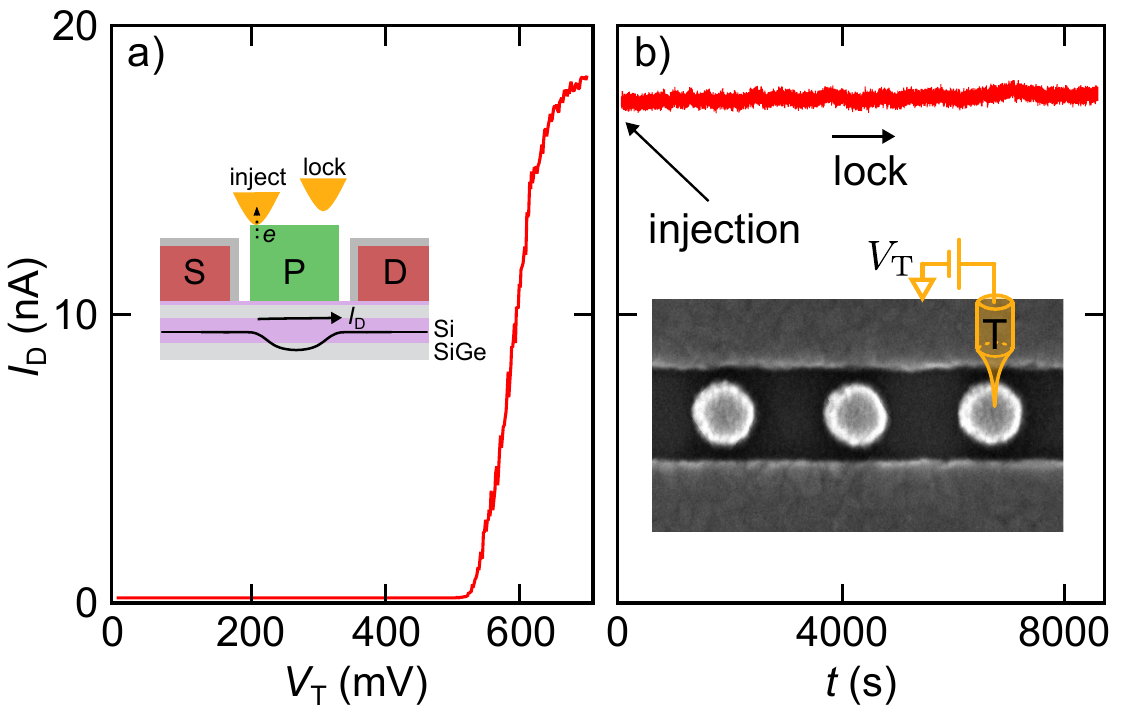}
	\caption{(a)~Single-layer device. The current through the device $I_{\rm D}$ measured as a function of the tip voltage $V_{\rm T}$ with the tip continuously touching the floating gate. The inset shows a cross-section of the device. (b)~$I_{\rm D}$ measured as a function of the time after the tip was withdrawn. Inset: SEM image of the device.}
	\label{fig_2_app}
\end{figure}

To cross-check our findings about the origin of the charge leakage in multilayer devices, we fabricated a single-layer device. Here, seven floating gates sit strictly on the Si/SiGe substrate, which now is the only path for charge to leak. Since all floating gates behave similarly, we present charge retention data from one of them, as shown in the inset of Fig.~\ref{fig_2_app}(b).

We start in the continuous injection mode sketched in Fig.~\ref{fig_2_app}(a), when the tip constantly touches the floating gate. As before, we measure the current $I_{\mathrm{D}}$ between the two ohmic contacts under the wide accumulation gates while applying a positive voltage $V_\mathrm{T}$ to the floating gate. The transistor-like turn-on curve shown in Fig.~\ref{fig_2_app}(a) lacks Coulomb oscillations due to the limited control of the confinement potential in single layer devices. Figure~\ref{fig_2_app}(a) shows the current measured as a function of time after the charge was locked with $V_\mathrm{T}=700~\mathrm{mV}$. In contrast to the multilayer device data presented above, we do not observe any charge leakage or tunneling events over several hours. As expected, the tunneling rate to the Si substrate is orders of magnitudes slower than the tunneling rate between overlapping gate layers.

\section{Conclusion and outlook}

In conclusion, we realized the ideal sample-and-hold circuit with a floating metallic gate fabricated on the surface of a Si/SiGe heterostucture. Utilizing the AFM tip as a switch allows us to reduce the plunger gate footprint down to $\sim$100 nm. The resulting stray capacitance of the floating gate is 2--3 orders of magnitude lower than in previous FET-based charge locking studies. The reduction of the stray capacitance allows us to probe much higher junction resistances through single electron counting than in previous studies. We find the average gate discharge tunneling rate to be of the order of one electron every few seconds in the overlapping gate architecture and multiple hours for single-layer devices. It follows that single-layer floating gates that do not require fast operation, such as those defining charge sensors, could be biased using the sample-and-hold approach, thereby significantly reducing the number of room temperature control lines per qubit. Looking forward, the AFM charging approach demonstrated here could be combined with tip-based dispersive readout~\cite{2202.10516}, enabling us to \textit{in-situ} tune and drag tip-induced QDs across the chip~\cite{doi:10.1063/1.5053756}.

\begin{acknowledgments}
Supported by Army Research Office grant W911NF-15-1-0149 and the Gordon and Betty Moore Foundation’s EPiQS Initiative through Grant No. GBMF4535. The authors thank V.~Khrapai, M.~F.~Gonzalez-Zalba and A.~Sachrajda for useful discussions. The authors thank F.~Borjans and A.~Mills for technical contributions to device fabrication and HRL Laboratories for providing the Si/SiGe heterostructures.
\end{acknowledgments}

\newpage
\bibliography{bib_PRL_chargelocking_v2}

\end{document}